\documentclass{article}
\usepackage{parskip}
\usepackage{microtype}
\usepackage{graphicx}
\usepackage{subfigure}
\usepackage{booktabs} 

\usepackage{hyperref}

\usepackage[accepted]{icml2024}

\usepackage{amsmath}
\usepackage{amssymb}
\usepackage{mathtools}
\usepackage{amsthm}

\usepackage[capitalize,noabbrev]{cleveref}

\theoremstyle{plain}

\theoremstyle{definition}

\theoremstyle{remark}

\usepackage[textsize=tiny]{todonotes}

\icmltitlerunning{Lost in Translation: The Algorithmic Gap Between LMs and the Brain}

\begin{document}

\twocolumn[
\icmltitle{Lost in Translation: The Algorithmic Gap Between LMs and the Brain}

\begin{icmlauthorlist}
\icmlauthor{Tommaso Tosato}{sch}
\icmlauthor{Pascal Jr Tikeng Notsawo}{sch2,sch3}
\icmlauthor{Saskia Helbling}{sch4}
\icmlauthor{Irina Rish}{sch2,sch3}
\icmlauthor{Guillaume Dumas}{sch,sch2}
\end{icmlauthorlist}

\icmlaffiliation{sch}{CHU Sainte Justine Research Center, Department of Psychiatry, Université de Montréal, Montreal, QC, Canada}\
\icmlaffiliation{sch2}{Mila, Montreal, QC, Canada}
\icmlaffiliation{sch3}{Department of Computer Science and Operation Research, Université de Montréal, Montréal, Canada}
\icmlaffiliation{sch4}{Ernst Strüngmann Institute for Neuroscience in Cooperation with Max Planck Society, Frankfurt am Main,
Germany}
\icmlcorrespondingauthor{Tommaso Tosato}{tosato.tommaso.office@gmail.com}

\icmlkeywords{Large Language Models, Neuroscience, Cognitive Science, Mechanistic Interpretability}

\vskip 0.3in
]

\printAffiliationsAndNotice{} 

\begin{abstract}
Language Models (LMs) have achieved impressive performance on various linguistic tasks, but their relationship to human language processing in the brain remains unclear. This paper examines the gaps and overlaps between LMs and the brain at different levels of analysis, emphasizing the importance of looking beyond input-output behavior to examine and compare the internal processes of these systems. We discuss how insights from neuroscience, such as sparsity, modularity, internal states, and interactive learning, can inform the development of more biologically plausible language models. Furthermore, we explore the role of scaling laws in bridging the gap between LMs and human cognition, highlighting the need for efficiency constraints analogous to those in biological systems. By developing LMs that more closely mimic brain function, we aim to advance both artificial intelligence and our understanding of human cognition.
\end{abstract}

\section{Introduction}
\label{introduction}
Large Language Models (LMs) have achieved remarkable performance on a wide range of language tasks, from machine translation to question answering \cite{brown2020language}. This success has led to speculation that LMs may provide valid models of human intelligence. However, comparisons between LMs and the brain are often limited to the input-output level, which can be misleading.

It would be incorrect to assume that LMs, because they produce human-like linguistic outputs, must process information similarly to the human brain. David Marr's levels of analysis framework \cite{marr1982vision} provides a valuable lens for comparing LMs and human language processing. Marr proposed that information processing systems can be understood at three distinct levels:

\vspace{-0.6em}
\begin{itemize}
\setlength\itemsep{0em}
\setlength{\itemsep}{0pt} 
\item \textbf{Computational level:} What is the goal of the computation? This level focuses on the abstract function being computed, independent of how it is implemented. In this context, it refers to tasks like language understanding, question answering, and translating between languages. Comparisons of input-output behaviors pertain to this level. 
\item \textbf{Algorithmic level:} How are the computational goals achieved? This level deals with the specific methods and strategies employed to transform inputs into outputs. In this context, it examines the mechanisms used to parse sentences, represent semantic information, and generate coherent answers. Analyzing which operations intermediate representations undergo pertains to this level.
\item \textbf{Implementational level:} How are these mechanisms physically realized? This level concerns the actual hardware (biological or artificial) that carries out the computations. For example, the brain uses spiking neurons and electrochemical signaling, while LMs run on silicon hardware with floating-point arithmetic.
\end{itemize}
\vspace{-0.6em}

Clarifying what aspects of cognition LMs are actually modeling is crucial. \citet{mahowald2024dissociating} argue that while LMs exhibit remarkable \textit{formal} linguistic competence (the ability to produce fluent and grammatical text), they still lack robust \textit{functional} competence (the capacity to use language to reason and achieve real-world goals). However, their performance has led to the idea that, beyond modeling language perception, they may also model more complex cognitive abilities such as reasoning \cite{jones2024ai}, theory of mind \cite{Strachan2024}, and building world models \cite{li2023structural}.

\citet{mitchell2023debate} contend that despite their seemingly intelligent behavior, LMs do not understand the data they process in the same way humans do. The brittleness and lack of robust generalization in these models are key indicators of their lack of true understanding. LMs are fundamentally retrieval systems that generate outputs by recognizing and interpolating patterns in their training data (i.e., \textit{approximate retrieval}) \cite{kambhampati2024can}. This leads to high performance for task instances which are well represented in the training data and failures for analogous instances which are less represented \cite{mccoy2023embers, wu2024reasoningrecitingexploringcapabilities, lewis2024using}. In contrast, human understanding is based on rich, causal mental models of the world, rather than massive statistical correlations learned by LMs. These mental models enable humans to make robust predictions, generalizations, and analogies, to reason compositionally and counterfactually, and to intervene in the world to test hypotheses actively.

A key challenge in understanding the true relationship between LMs and the brain is to move beyond comparisons of input-output behavior and intermediate representations (i.e., the \textit{computational level}), and focus on probing the internal processes of these systems (i.e., the \textit{algorithmic level}) using causal interventions and mechanistic interpretability. This argument is developed in the first part of the paper. We then proceed to identify key properties in the architectures and training procedures of LMs which differ from the brain and propose possibilities to reconcile them. By doing so, we aim to develop LMs that better capture human cognitive processing, and as such, could serve as more faithful models of the brain.

\section{Probing Representations and Processes}

To determine whether LMs process information in ways analogous to the human brain, it is essential to look beyond surface-level behavior. While LLMs have recently been claimed to pass the Turing test \cite{jones2024peopledistinguishgpt4human}, \citet{dumas2014human} argue that the classic Turing test, which focuses solely on external behavior, is insufficient for investigating the genuineness of the internal mechanisms of machine intelligence. This is especially true for language, where the ELIZA effect can lead people to attribute human-like qualities to any system capable of generating coherent text \cite{weizenbaum1976computer}. Moreover, performance comparisons between LMs and humans at the behavioral level are subject to several pitfalls. These include models trained specifically to solve benchmarks that claim to quantify general cognitive abilities but do so in arbitrary and limited ways \cite{liu2024ecbd}, and the risk of data contamination, where test data may have been included in the model's training set. This section outlines possible approaches for comparing LMs and the brain beyond the input-output level:

\subparagraph{Hierarchical Correspondences of Representations.}
Research in computer vision has shown that representations learned by Convolutional Neural Networks across layers correlate with brain activity in different regions of the visual hierarchy \cite{yamins2016using, kriegeskorte2015deep, khaligh2014deep}. This was demonstrated by using techniques like representational similarity analysis (RSA) \cite{kriegeskorte2008representational} to compare the geometry of neural representations across models and brain regions.
Similarly, in the language domain, studies have found correspondences between LM activations across layers and brain activity in different parts of the language network \cite{caucheteux2023evidence, millet2023realistic}. Shared representational spaces have been uncovered through \textit{encoding models}, which produce simulated brain activity from LM embeddings through linear regression and correlate the simulated activity with recorded neural data \cite{goldstein2022shared}. The correlation strength is reflected in an index called \textit{brain score} \cite{schrimpf2021neural}. However, it is important to note that both the visual system and the language network are not simple feedforward hierarchies but rather highly recurrent networks with pervasive feedback connections and extensive cross-talk between regions \cite{felleman1991distributed,pessoa2023entangled}.
Moreover, it is crucial to understand that representational similarity between LMs and brains pertains primarily to the computational level, as it captures the input-output mappings of intermediate computational steps, but does not necessarily imply that similar operations or algorithms are used to transform those representations \cite{schyns2022degrees,antonello2024predictive,tuckute2024language}. Two systems can exhibit similar representations while differing substantially in their algorithms. This is evidenced by the fact that not only causal transformers (e.g., GPT), but also masked transformers (e.g., BERT), LSTMs and RNNs generate internal representations that correlate with brain activity, and there is no clear agreement on which one correlates best \cite{pasquiou2022neural, anderson2021deep, toneva2019interpreting, oota2022long}. To make claims about algorithmic similarities, we need evaluation approaches that can probe the internal dynamics and causal structure of LMs \cite{belinkov2022probing} and compare them with results obtained using similar approaches in the brain.

\subparagraph{Mechanistic Interpretability.} This approach involves analyzing and understanding the internal components and processes of a machine learning model to explain how it transforms inputs into outputs at a granular, algorithmic level \citep{elhage2021mathematical}. It can be used to find units or circuits within LMs that encode particular syntactic or semantic properties. However, individual neurons in LMs respond to a variety of seemingly unrelated and heterogeneous features, posing a challenge for interpreting the network's behavior. Recent work addresses this issue by using sparse autoencoders to extract monosemantic features from transformer language models \cite{bricken2023towards,cunningham2023sparse,templeton2024scaling}, including abstract concepts that generalize across modalities and languages. These learned features are significantly more interpretable than the model's original neurons and exhibit interesting properties such as universality across different random seeds and feature splitting as the number of learned features increases. Comparing these features to those recorded in brain networks \cite{jamali2024semantic} could provide a more accurate correspondence between LMs and the brain. Notably, neurons in the brain are also semantically heterogeneous to a certain degree \cite{leonard2024large, tye2024mixed}, suggesting that some degree of polysemanticity may be a general property of artificial and biological neural systems.

\subparagraph{Ablation and Stimulation Studies.} These techniques can provide causal insights in both the brain and LMs. Lesions \cite{broca1861remarks,dronkers2004lesion} and electric stimulation \cite{doron2015single, rofes2019language} studies have allowed mapping of the functional organization of language in the human brain. More recently, optogenetics has enabled targeting and re-activating specific neuronal ensembles, which were previously activated by certain stimuli \cite{liu2012optogenetic}. 
Analogous experiments in LMs, such as pruning, editing or fixing specific weights, can reveal how different parts of the network contribute to linguistic behavior \cite{meng2023locating,vig2020investigating}. Additionally, activating or suppressing specific extracted features can influence the model's output, and allow steering it in specific directions \cite{templeton2024scaling, marks2024sparsefeaturecircuitsdiscovering}. For instance, clamping certain features to high values can induce models to generate specific types of content or alter their behavior in predictable ways. Comparable effects of perturbations in LMs and the brain would suggest algorithmic equivalence.

\subparagraph{Controlled Linguistic Probes and Interventions.} These can be combined with interpretability techniques to gain a more mechanistic understanding of language processing in LMs \cite{marvin2018targeted, geiger2021causal, zhou2024constructions}. By systematically perturbing inputs or fine-tuning LMs on specific tasks, we can observe how internal representations change in response to these interventions. Similar approaches in neuroscience can provide complementary insights and offer ground for comparison.

\subparagraph{Meta-representations as Representation of Processes}. Proposed by \citet{kanai2024meta}, this framework offers a promising approach for comparing the internal workings of LMs and the brain at the algorithmic level by focusing on the processes that generate representations, rather than just the representations themselves. In the case of LMs, we can use the patterns of weights across different layers and attention heads as a way to represent the processes that generate the model's outputs. We can then compare the two at the algorithmic level by relating these weight patterns to process models of the brain \cite{dohmatob2020dark}, based on, e.g., structural and functional connectivity.

\section{Toward Brain-Inspired Language Models}

While Language Models (LMs) have achieved human-like performance on several tasks, they still differ significantly from the human brain in the ways they achieve these results. As discussed in the previous section, better methods to test for these differences at the algorithmic level are needed. In this section, we explore how LMs and the brain differ in terms of architectures and learning dynamics, and how these differences lead to divergences at the algorithmic level. By examining these disparities, we can identify key areas where LMs could be modified to more closely resemble the brain's information processing mechanisms. The motivation for reconciling these differences is twofold: 1) By developing LMs that more closely mimic the brain's architecture and dynamics, we can provide neuroscientists with more accurate computational models of language processing in the brain. This can lead to new hypotheses and insights in neuroscience \cite{jain2024computational}, fostering a deeper understanding of human cognition. 2) Incorporating brain-inspired features into LMs may result in artificial systems that exhibit more human-like intelligence. This could lead to AI that is more flexible, efficient, and capable of handling reasoning and planning in ways that current models struggle with. 

\subparagraph{Sparse Connectivity and Modularity.}
The brain is characterized by sparse connectivity and modularity, with specialized networks for different cognitive functions \cite{bassett2018nature, seguin2023brain}. In contrast, transformer models have dense, all-to-all connectivity. During development, the brain undergoes synaptic pruning, resulting in sparser, more efficient networks \cite{paolicelli2011synaptic}. Similar pruning emerges in biologically inspired spiking neural networks \cite{volzhenin2022multilevel}.
To incorporate some level of modularity in LMs, Mixture-of-experts (MoE) architectures \cite{xue2024openmoe} can be employed, leading to improved computational efficiency and  more functionally specialized sub-networks. However, the modules in MoE are not interconnected, hierarchically organized, or able to learn to attend to specific parts of the inputs. The Recurrent Independent Mechanisms (RIMs) architecture \cite{goyal2019recurrent} addresses these limitations by introducing sparse interactions among functionally specialized modules that interact sparsely through attention. This enables these modules to learn to attend to specific input parts and allows for hierarchical stacking.
Sparsity can be promoted through adjustment of attention mechanisms (e.g., limiting the number of tokens each token attends to \cite{child2019generating}), choosing regularization methods alternative to L2 (e.g., L1 regularization and elastic net), and applying dropout and post-training pruning. 

\subparagraph{Internal States.}
The human brain maintains and constantly updates rich internal representations over time, whereas LMs have a fixed context window, and only feed-forward connections. Incorporating internal states into LMs could change the way they maintain and use context. Recent work on adding recurrence to transformer models \cite{hwang2024transformerfam} is a step in this direction. State space models \cite{gu2023mamba}, which explicitly model the dynamics of latent states, are another promising approach. Models that learn to update and maintain internal representations, may also dynamically encode and integrate information over time in a brain-like manner \cite{hasson2015hierarchical}. Moreover, recurrent processing appears to be critical for consciousness \cite{chalmers2023largelanguagemodelconscious}.

\subparagraph{Learning Dynamics.}
The brain's learning process is characterized by continual, incremental acquisition of knowledge over a lifetime, utilizing moderate resources efficiently. A key feature of brain development is the presence of critical periods, during which the brain is especially receptive to specific types of input and learning experiences \cite{hensch2005critical}. Moreover, brain learning dynamics maintain a delicate balance between acquiring new information and selectively forgetting unnecessary details.
In contrast, current LMs are typically trained from scratch using random weight initialization on massive corpora, requiring multiple passes over large datasets and consuming enormous computational resources. Additionally, fine-tuning or training LMs on new tasks often leads to catastrophic forgetting \citep{kirkpatrick2017overcoming}, highlighting the lack of balance in retaining old knowledge while acquiring new information.
To enable more brain-like learning in LMs, several approaches can be considered: (1) \textit{Continual Learning}: Adapting continual learning techniques such as elastic weight consolidation \cite{kirkpatrick2017overcoming} and experience replay \cite{rolnick2019experience} could help LMs maintain a balance between learning new information and preserving important past knowledge. (2) \textit{Curriculum Learning:} Gradually increasing the difficulty of training data \citep{10.1145/1553374.1553380} can enhance learning efficiency. The PAIRED algorithm \cite{dennis2020emergent}, which uses a multi-agent setup where an adversary generates increasingly challenging environments for the learning agent, is a possible approach to implementing adaptive curriculum learning. (3) \textit{Critical Periods:} Implementing a system of "sensitive periods" in LM training could mimic the brain's critical periods. This could involve dynamically adjusting learning rates for different parts of the network during training. For instance, early in training, the model could have higher plasticity in lower layers to learn basic linguistic features, gradually shifting focus to higher-level abstractions in later stages. Interestingly, current artificial neural networks already show some similarities to biological learning, exhibiting progressive learning of features of increasing complexity \cite{nakkiran2019sgd, mangalam2019deep}. 

\subparagraph{Embodiment, Grounding, and Active Learning.}
Humans learn language through situated interactions with the physical and social world, grounding linguistic meanings in perceptual, motor, and affective experiences \cite{bisk2020experience, barsalou2008grounded, di2018linguistic}. LMs, in contrast, are typically trained on disembodied text data with a fixed training objective, limiting their ability to capture the full depth of linguistic meaning. To address this limitation, multimodal models that integrate information from vision, audition, and other sensory modalities are a promising approach \cite{chameleonteam2024chameleon}. However, true embodiment may require more than just passive perception—it may need active interaction and exploration \cite{engel2013s}. Coupling LMs with external execution modules that can perform actions or specialized reasoning tasks could provide an action space. However, just adding an action space when the model's weights are frozen won't generate active learning. For true active learning, the model needs to be able to select the training data through actions. Integrating LMs into robotic systems \cite{tellex2020robots, embodimentcollaboration2024openxembodimentroboticlearning} is another avenue for grounding language in embodied experience.

\subparagraph{Social Skills and Interactive Learning.}
The brain's language system is fundamentally geared towards communication and social interaction, with dedicated neural mechanisms for cognitive skills like pragmatics, social reasoning, and theory of mind \cite{scott2019speech, hagoort2014neurobiology}. While recent studies have shown that LMs are starting to exhibit some of these abilities \cite{Strachan2024}, it is unclear if they are achieving these capabilities in a way related to humans. To improve LMs' social abilities, they could be trained on socially interactive data, such as multi-turn dialogues, incorporating explicit representations of communicative intentions and the mental states of interlocutors. Training objectives that maximize relevance for a communicative context or reward successful and attuned communication in interactive settings \cite{jaques2019social} could lead to more human-like language use. Multi-agent setups where LMs interact with each other or humans in goal-directed dialogues are promising \cite{bolotta2022social}. By engaging in cooperative or competitive language games, LMs could learn to reason about the beliefs and intentions of other agents. Inverse reinforcement learning techniques \cite{hadfield2016cooperative} could help the LMs to infer and align with the communicative goals of human interlocutors.

\subparagraph{Reasoning and Compositional Representations.}
LMs' ability to perform reasoning and planning remains limited compared to humans \cite{wu2023reasoning,valmeekam2023planbench, zhou2023algorithms}. Techniques like chain-of-thought prompting \cite{wei2023chainofthought} and the use of scratchpads \cite{nye2021show} to generate step-by-step reasoning traces, have shown promise in improving reasoning abilities. Using reasoning traces as fine-tuning data has also been shown to enhance performance \cite{NEURIPS2022_639a9a17}. However, the brain's ability to perform logical reasoning likely relies on more structured, compositional representations of meaning \cite{dehaene2015neural, lake2017building}.
GFlowNets \cite{bengio2021flow} are generative models that can sample from complex joint distributions, such as the distribution over sentences and their meanings. GFlowNets could potentially capture the brain's ability to generate diverse outputs and learn structured, compositional representations. Initial applications of GFlowNets to LMs are already showing promising results \cite{hu2024amortizing}.
Another idea is to move beyond next-word prediction as the primary training objective for LMs. The brain does not simply predict the next word in a sequence but constructs rich, hierarchical meaning representations that span multiple words and sentences \cite{heilbron2022hierarchy, lerner2011topographic}. Training objectives that encourage LMs to predict larger chunks of text, such as entire sentences or paragraphs, could lead to more structured and compositional representations \cite{gloeckle2024better}.

\subparagraph{Oscillatory Dynamics.} One key feature of language processing in the brain is the presence of oscillatory neural dynamics at multiple timescales, with different frequency bands appearing to track the structure of language at different levels, from phonemes to words to phrases \cite{giraud2012cortical, ding2016cortical, gross2013speech}. These oscillations are thought to play a crucial role in segmenting and chunking continuous speech into discrete units.
Incorporating oscillatory mechanisms in deep learning architectures, especially when the input is raw auditory signals, could potentially improve their efficiency and lead to more human-like speech processing, given the existing gaps \cite{tuckute2023many}. Simulation work has been exploring substituting standard neural network nodes with damped harmonic oscillators \cite{rusch2020coupled, effenberger2022functional}, demonstrating the potential use of oscillatory dynamics for computation.

\subparagraph{Allometry and Scaling Laws.}
Allometry studies the relationship between the size and shape of biological systems. Scaling laws in the human brain, such as the relationship between brain size and neuron count, play a crucial role in the evolution of cognitive abilities, including language \cite{changeux2021connectomic}. Recently, the field of neural scaling laws has similarly started to chart the relationships between the number of parameters, dataset size, computing cost, and performance of machine learning models \cite{kaplan2020scaling}. These scaling relationships can be seen as global signatures of complex systems \cite{west2018scale}, providing a quantitative way to compare brains and LMs. However, LMs benefit from virtually unlimited computational resources, leading to overparameterization without the same efficiency constraints as biological systems \cite{west1999origin}. To promote algorithmic similarity, it may be essential to introduce analogous pressures on LMs. Implementing such constraints can help bridge the gap between the computational prowess of LMs and the adaptive efficiency of biological neural systems, ensuring more frugal and meaningful application of scaling laws in LM development.
\section{Conclusion}

The rapid advancement of Language Models (LMs) has led to impressive performance on various linguistic tasks, prompting comparisons with human cognition. However, this paper argues that despite achieving human-level performance in several tasks (computational level equivalence), LMs exhibit significant divergence from human cognition in how these performances are achieved (algorithmic level discrepancy). We contend that the intelligence emerging from current LMs is fundamentally different from human cognition. As models continue to scale, this divergence may widen, potentially resulting in AI systems that excel at specific tasks but lack the hallmark characteristics of human intelligence: flexibility, robustness, and generalization capabilities. 

In this paper, we first examined approaches for comparing LMs and the brain beyond input-output level assessments. We highlighted the limitations of solely analyzing intermediate representations and emphasized the need to understand the processes driving transformations between these representations. We then explored major differences in architecture and training dynamics between LMs and biological neural networks that likely contribute to algorithmic-level divergence, and we proposed several neuroscience-inspired approaches to narrow this divide. 

While these biologically-inspired approaches may not immediately enhance computational efficiency or task performance, they are crucial for developing LMs that can serve as more accurate models of human cognition. Such models would offer in silico representations of cognitive processes to neuroscientists and psychologists, illuminating key mechanisms underlying human-like intelligence. Additionally, by constraining the space of possible algorithms to those that are more biologically plausible, we may discover novel architectures and learning paradigms that capture the key principles of human cognition, and lead to AI systems that exhibit the adaptability and generalization capabilities characteristic of human intelligence.

\bibliography{paper}
\bibliographystyle{icml2024}

\end{document}